 \documentclass[final,5p,times,twocolumn,authoryear]{elsarticle}
\usepackage{graphicx}

\usepackage{amssymb}
\usepackage{lipsum}
\usepackage{xcolor}

\newcommand{\ignore}[1]{}

\journal{Physics Letters B}

\title{A Quantum Computational Perspective on Spread Complexity}
\author[1]{Cameron Beetar}
\author[2]{Eric L Graef}
\author[1]{Jeff Murugan}
\author[2]{Horatiu Nastase}
\author[1]{Hendrik J R Van Zyl,}
\affiliation[1]{organization={The Laboratory for Quantum Gravity \& Strings, Department of Mathematics and Applied Mathematics, University of Cape Town, Cape Town, South Africa}}
\affiliation[2]{organization={Instituto de F{i}sica Te{o}rica, UNESP-Universidade Estadual Paulista},
            addressline={R. Dr. Bento T. Ferraz 271, Bl. II}, 
            city={Sao Paulo},
            postcode={01140-070, SP}, 
            country={Brazil}}

\begin{document}

\begin{abstract}
    We establish a direct connection between spread complexity and quantum circuit complexity by demonstrating that spread complexity emerges as a limiting case of a circuit complexity framework built from two fundamental operations: time-evolution and superposition. Our approach leverages a computational setup where unitary gates and beam-splitting operations generate target states, with the minimal cost of synthesis yielding a complexity measure that converges to spread complexity in the infinitesimal time-evolution limit. This perspective not only provides a physical interpretation of spread complexity but also offers computational advantages, particularly in scenarios where traditional methods like the Lanczos algorithm fail. We illustrate our framework with an explicit $SU(2)$ example and discuss broader applications, including cases where return amplitudes are non-perturbative or divergent.
\end{abstract}

\maketitle

\section{Introduction}
Quantum complexity has emerged as a unifying theme across diverse areas of physics, from the study of black hole dynamics and holography to quantum chaos and quantum computing. In high-energy physics, complexity measures have been proposed to probe the growth of operators under Heisenberg evolution (\cite{Parker:2018yvk}), the emergence of spacetime geometry (\cite{Pedraza:2022dqi}), and the onset of scrambling in strongly interacting systems (\cite{Bhattacharjee:2022vlt}). Meanwhile, in quantum information science, circuit complexity quantifies the minimal resources required to prepare a target state from a reference state, offering insights into the intrinsic hardness of quantum simulation and computation (\cite{Nielsen:2012yss}, \cite{Jefferson:2017sdb}). Despite their shared conceptual roots, these two perspectives—dynamical complexity in physical systems and operational complexity in quantum circuits—have largely evolved in parallel (\cite{Chapman:2021jbh}).

This work bridges this divide by establishing a concrete connection between {\it spread complexity} (\cite{Balasubramanian:2022tpr}), an increasingly popular measure of state complexity in quantum dynamics, and {\it circuit complexity}, a fundamental notion in quantum information (\cite{Nielsen:2012yss}). Spread complexity, defined through the Krylov basis generated by Hamiltonian evolution (\cite{Parker:2018yvk}), has proven effective in diagnosing quantum chaos (\cite{Dymarsky:2019elm}, \cite{Bhattacharyya:2019txx}, \cite{Avdoshkin:2022xuw}), integrability-breaking transitions (\cite{Balasubramanian:2024ghv}, \cite{Camargo:2024deu}), operator growth (\cite{Parker:2018yvk}), topological phase transitions (\cite{Caputa:2022eye}, \cite{Caputa:2022yju}), PT-symmetric transitions (\cite{Beetar:2023mfn}, \cite{Bhattacharya:2024hto}) and many other dynamical features. However, its physical interpretation has remained abstract, and its computation often relies on the Lanczos algorithm (\cite{Lanczos1950AnIM}), which can fail for systems with divergent moments or non-perturbative dynamics. Here, we show that spread complexity arises naturally as the minimal cost of synthesizing a target state using a simple quantum circuit composed of two fundamental operations: time-evolution (a unitary gate) and superposition (a beam-splitting operation and beam-combining operation). By assigning computational costs judiciously, we demonstrate that the optimal circuit reproduces the Krylov basis in the limit of infinitesimal time steps, thereby linking spread complexity directly to a physically intuitive quantum information protocol.

Our framework not only clarifies the operational meaning of spread complexity but also offers practical advantages. Traditional methods for computing spread complexity require derivatives of the return amplitude or high-order moments of the Hamiltonian (\cite{Muck:2022xfc}, \cite{Nandy:2024htc}), which may diverge or be inaccessible in certain systems. In contrast, our approach relies solely on discrete evaluations of the return amplitude, making it robust in non-perturbative regimes. We illustrate this with an explicit example in an $SU(2)$ system, where the circuit complexity converges to the known spread complexity as the time step vanishes. Moreover, our method generalizes naturally to multi-operator settings, suggesting a pathway to extend the notion of Krylov complexity to more intricate quantum circuits.

This work situates spread complexity within the broader landscape of quantum information and high-energy physics, revealing it as a special case of a more general circuit complexity measure. By doing so, it opens new avenues for cross-pollination between these fields—for instance, in designing quantum algorithms to simulate complex dynamics or in leveraging physical intuition from many-body systems to optimize quantum circuits. Our results also invite future explorations of complexity in open quantum systems, field theories, and holographic settings, where the interplay between unitary and non-unitary operations could yield new insights into the nature of quantum complexity. By anchoring spread complexity in a quantum circuit model, we hope that this work not only demystifies an abstract measure but also enriches the toolkit for studying complexity across quantum physics.

\section{Computational Setup}

\subsection{Target space with one unitary and superpositions}

\label{CircuitComplexitySection}

The setup we will consider is one where target states may be generated by means of three operations.  The quantum states will be treated as spread out over a number of $n$  ``beams" so that the Hilbert space of the total system consists of vectors of the form
$$    \left( \begin{array}{c}  |\psi_1\rangle \\ |\psi_2\rangle \\ \vdots \\ |\psi_n\rangle  \end{array} \right)\ , \ \ \ \textnormal{where} \ \ \ \sum_{i=1}^n \langle \psi_i|\psi_i\rangle = 1 \space.$$   Our reference and target states are taken to be confined to a single ``beam''.  In between, however, the state may manipulated by splitting and recombining channels similar to beams in a photonics setup or particles in a multi-slit experiment.  The first operation we thus consider is a quantum beam splitter which maps an $n$-beam vector to an $(n+1)$-beam vector as\footnote{Though it may appear so at first glance, this operation does not violate the no-cloning theorem.  Indeed the beam splitting may be realised as a unitary operation (along with more precise notation) - see \ref{Experiment}.}
\begin{equation}
Q_k( z ) \left( \begin{array}{c}   \vdots \\ |\psi_k\rangle \\  \vdots  \end{array}   \right) = \left( \begin{array}{c}  \vdots \\ \frac{1}{\sqrt{1 + |z|^2}} |\psi_k\rangle  \\  \frac{z}{\sqrt{1 + |z|^2}} |\psi_k\rangle  \\ \vdots  \end{array}\right) \ ,
\end{equation}
where we have acted on the $k$'th beam in the setup.  The second operation we will consider is to recombine two beams. 
 This is a superposition which maps an $(n+1)$-beam vector to an $n$-beam vector as 
\begin{eqnarray}
S_{kl} \left( \begin{array}{c } \vdots \\ |\psi_k\rangle  \\ \vdots \\ |\psi_l\rangle \\ \vdots  \end{array} \right) & = & \left( \begin{array}{c } \vdots \\ N(|\psi_k\rangle  + |\psi_l\rangle) \\ \vdots \\ \vdots  \end{array} \right) \ ,  \\
N &=& \sqrt{ \frac{\langle \psi_k| \psi_k\rangle + \langle \psi_l | \psi_l\rangle}{ ( \langle \psi_k| + \langle \psi_l|)( | \psi_k \rangle +  | \psi_l \rangle)   }    } \ ,\nonumber
\end{eqnarray}
where we have combined beams $k$ and $l$ in a way that preserves the total probability.  The final operation we will consider is a single unitary gate which can act on any beam in the setup e.g.
\begin{equation}
A_k  \left( \begin{array}{c } \vdots \\ |\psi_k\rangle  \\   \vdots \end{array} \right) =  \left( \begin{array}{c } \vdots \\ A|\psi_k\rangle \\ \vdots  \end{array} \right) \ \ \ \textnormal{with} \ \ \ A^\dag A = I \ .
\end{equation} 
The space of target states (in a single beam) may be organised into the sets 
\begin{equation}
S_k = \left\{ |\psi\rangle =  \sum_{j=0}^k \alpha_j A^j \ignore{|\tilde{K}_0\rangle} |\kappa_0\rangle  \ \ \  ; \ \  \langle \psi | \psi \rangle  = 1 \right\} \ ,   \label{targetStatesk}
\end{equation}
involving up to $k$ applications of the unitary gate $A$.  Of imminent relevance is the set of orthogonal states\footnote{Note that we use round brackets here for the bra and ket.  This is to distinguish states that are not necessarily normalised from those that are.} 
\begin{eqnarray}
|\kappa_{n}) & = & \sum_{m=0}^n c_{n,m} A^{m} |\ignore{\tilde{K}_0}\kappa_0\rangle \ , \nonumber \\
(\kappa_{n}| & = & \sum_{m=0}^n c^*_{n,m}  \langle \ignore{\tilde{K}_0}\kappa_0|(A^\dag)^{m} \ ,  \nonumber \\
(\kappa_a| \kappa_{b}) & = & \delta_{a,b} (N_a)^2 \ ,\label{KtildeBasis}
\end{eqnarray}
which provides an orthogonal basis for the space of target states.   

\subsection{Target state synthesis}

An important feature of the computational setup described above is that a desired target state may be generated in many different ways.  For example, it is a mathematically trivial step to write 
\begin{equation}
A |\kappa_0\rangle = \left(  A|\kappa_0\rangle + \alpha |\kappa_0\rangle   \right) - \alpha |\kappa_0\rangle \ . \label{splitBeam}
\end{equation}
Operationally, however, the above is non-trivial and would require a combination of the operations $Q, S$, and $A$.  We have sketched one such example pictorially in Fig. (\ref{fig:C1Design}), but this is far from the only possibility.  \\ \\
For our purposes, it will be useful to combine the operations into channels (for details of their design, we refer the reader to \ref{Experiment}) that satisfy $$ \langle \kappa_0| \hat{C}^\dagger_{k} \hat{C}_k |\kappa_0\rangle = 1.$$  Relevant for the space $S_1$ for example (\ref{targetStatesk}), we can combine $Q, S$ and a single operation of $A$ into a channel 

\begin{eqnarray}
\hat{C}_1(\alpha_{1, 0}, \alpha_{1, 1})  & = &    \frac{\left( \alpha_{1, 0} |\kappa_0\rangle + \alpha_{1, 1}  |\kappa_1\rangle  \right)}{\sqrt{|\alpha_{1, 0}|^2 + |\alpha_{1, 1}|^2}} \langle \kappa_0 |\ .  \label{eq:C1Design}
\end{eqnarray}
Similarly (relevant for $S_2$), we can can combine $Q$, $S$, and two applications of $A$ into a channel
\begin{eqnarray}
\hat{C}_2(\alpha_{2, 0}, \alpha_{2, 1}, \alpha_{2, 2}) & = & \frac{\left( \alpha_{2, 0} |\kappa_0 \rangle + \alpha_{2, 1} |\kappa_1 \rangle + \alpha_{2, 2} |\kappa_2 \rangle \right)}{\sqrt{|\alpha_{2, 0}|^2 + |\alpha_{2, 1}|^2 + |\alpha_{2, 2}|^2}}\langle \kappa_0|  \ . \nonumber  \\ & & \label{C2Design}
\end{eqnarray}
Focusing on the first channel, we may synthesise a state from $S_1$ as 
\begin{equation}
\gamma_1 |\kappa_1\rangle + \gamma_0 |\kappa_0 \rangle = c_1 \hat{C}_1 |\kappa_0\rangle + c_0 |\kappa_0\rangle   \ ,\label{channel1Synth}
\end{equation}
which requires
\begin{eqnarray}
c_1 & = & \frac{\sqrt{|\alpha_{1, 0}|^2 + |\alpha_{1, 1}|^2}}{\alpha_{1, 1} } \gamma_1 \ ,  \nonumber \\
c_0 & = & \frac{\alpha_{1, 1} \gamma_0 - \alpha_{1, 0} \gamma_1}{\alpha_{1, 1}} \ .
\end{eqnarray}
Using the first and second channel we may synthesise a state from $S_2$ as 
\begin{equation}
\gamma_2' |\kappa_2\rangle + \gamma_1' |\kappa_1 \rangle + \gamma_0' |\kappa_0\rangle = \left( c_2' \hat{C}_2 + c_1' \hat{C}_1  + c_0' \right) |\kappa_0\rangle \ ,  \label{channel2Synth}
\end{equation}
which requires
\begin{eqnarray}
c_2' & = & \frac{\sqrt{|\alpha_{2, 0}|^2 + |\alpha_{2, 1}|^2 + |\alpha_{2, 2}|^2}}{\alpha_{2, 2}}\gamma_2'  \ , \nonumber \\
c_1' & = & \frac{\sqrt{|\alpha_{1, 0}|^2 + |\alpha_{1, 1}|^2}}{\alpha_{1, 1} \alpha_{2, 2}}(\alpha_{2, 2} \gamma_1' - \alpha_{2, 1} \gamma_2') \ , \nonumber \\
c_0' & = & \frac{\alpha_{1, 0}(\alpha_{2, 2} \gamma_1' - \alpha_{2, 1} \gamma_2') + \alpha_{1, 1}(\alpha_{2, 0} \gamma_2' - \alpha_{1, 1} \gamma_0')}{\alpha_{1, 1} \alpha_{2, 2}} \ .
\end{eqnarray}

Clearly, we can design many different channels that may produce a desired target state.  Furthermore, note that all the coefficients $c_{j}'$ must satisfy $|c_j'| \leq 1$ in order for the computation to be physical i.e. for the state to be producible by the set of gates used on the right-hand side.

\subsection{Computational Cost}

Each different algorithm for synthesizing a desired target state may have a different computational cost.  We need to decide how to assign computational cost to the above.  In what follows (and what will ultimately make contact with spread complexity) we will make the following choices: 
\begin{enumerate}
\item We will take beam splitting and superposition to have \textbf{no cost}.  This is, of course, a major theoretical simplification but one that is necessary to make contact with state spread complexity.  It may be justified physically if the cost of applying the unitary gate $A$ is far greater than the operations $Q$ and $S$.  In quantum computational frameworks this seems reasonable, since the Hadamard gate implements superpositions whereas a general unitary $A$ may in fact be composed of the product of many elementary operations.  We  will normalise the cost of applying gate $A$ to be $1$. 
\end{enumerate}
In future work it would be interesting to consider the impact of assigning a cost to the operators $Q$ and $S$.  We expect that will impact the optimal channel designs in a significant way.  To see this intuitively, consider the opposite limit, where implementing the unitary $A$ is far cheaper than performing superpositions.  In this case the family of states $\left\{  A^n |\kappa_0\rangle  \right\}$ have a negligible cost compared to any state composed from taking superpositions.  The setup then more resembles that of multi-seed Krylov complexity, see \cite{Craps:2024suj}, with a family of states making up the reference set, as opposed to a single reference state.
\begin{enumerate}
\setcounter{enumi}{1}
\item When synthesizing a target state through the use of channels such as (\ref{channel1Synth}) we will take the cost of the channel to be the number of applications of $A$, i.e. channel $\hat{C}_k$ has cost $k$.  This is the usual notion of circuit depth complexity (for the gate $A$) in our setup. 
\end{enumerate}
It immediately follows from the considerations above that the state $|\kappa_n\rangle$ will have a cost of at least $n$ since it can only start appearing at a circuit depth of $n$.  \\ \\
Finally, we need to decide how to assign a cost to a state made up from the sum of different channels. From the circuit depth perspective a state with non-vanishing overlap with $|\kappa_n
\rangle$ should be assigned a complexity of at least $n$.   However, though our target state may require the implementation of channel $\hat{C}_k$, its contribution may only be a small fraction of the target state.  This is a direct consequence of allowing a zero-cost superposition operation.  
\begin{enumerate}
\setcounter{enumi}{2}
\vspace{0.13cm}
\item  A natural choice of cost function would be to weigh the probability that (in its preparation) the target state passed through channel $C_k$ by the computational cost $k$.  A simpler\footnote{This cost function does not quite give the expected cost of producing the target state $|\psi\rangle$ since the output of the channels are not orthogonal.  After optimizing, this will, however, be the correct interpretation.} (but closely related) choice of cost for an arbitrary state $|\psi\rangle$ is 
\begin{equation}
\sum_{k}\sum_{m} \ k |\langle \psi| \kappa_m\rangle \langle \kappa_m| \hat{C}_k |\kappa_0\rangle|^2.  \label{costFunc}
\end{equation}
which we will be using in what follows.  
\end{enumerate}
Having now chosen the computational cost we now turn to the computational complexity of a desired target state.  This is defined as the minimal computational cost required to synthesise the state.  In our setup this means minimizing the cost function over the channel design parameters (e.g. the $\alpha_{1, i}$ in (\ref{eq:C1Design}) and $\alpha_{2,i}$ in (\ref{C2Design})).   We find that the optimal channels are
\begin{eqnarray}
\hat{C}_1 & = & |\kappa_1\rangle\langle \kappa_0|\ ,    \nonumber \\
\hat{C}_2 & = & |\kappa_2 \rangle \langle \kappa_0| \ , \nonumber \\ 
& \vdots &     \nonumber \\
\hat{C}_n & = & |\kappa_n \rangle \langle \kappa_0|\ . \nonumber
\end{eqnarray}
See \ref{AppendixOptimal} for details. The optimal channels thus produce mutually orthogonal states.  The physical intuition behind this is clear (and in essence identical to \cite{Balasubramanian:2022tpr}).  The $|\kappa_m\rangle$ component of a target state can be produced by channel $\hat{C}_m$ at cost proportional to $m$ or by a more expensive channel.  The optimal algorithm produces as much of the $|\kappa_m\rangle$ component at the minimal cost as is possible.  This leads directly to subsequent channels giving an output that is orthogonal to the output of channel $\hat{C}_m$.  \\ \\
The circuit complexity for a target state $|\psi\rangle$ may be written as the expectation value
$$ \langle \psi| \hat{K} |\psi \rangle\ , \ \ \ \textnormal{where} \ \ \ \hat{K} = \sum_{n} n | \kappa_n\rangle \langle\kappa_n| \ . $$

\section{Connection to spread complexity}

The computational cost we have computed above may remind the reader of spread complexity (\cite{Balasubramanian:2022tpr}) where a Krylov basis (\cite{Parker:2018yvk, Caputa:2021sib}) is generated by repeated action of a system Hamiltonian on a specified reference state.  There is, however, one key difference: the Hamiltonian that generates the Krylov basis is (typically) a hermitian operator.  In contrast, the gate $A$ we have used above is a unitary operator.  We stress that the basis $|\kappa_n\rangle$ is not the Krylov basis, $|K_n\rangle$.  In particular, the unitary gate expressed in the basis (\ref{KtildeBasis}) is {\it not tri-diagonal} unlike the Hamiltonian expressed in the Krylov basis.  
\\ \\
The two approaches are, however, closely related and, indeed, the usual spread complexity may be recovered if our unitary implements a small time-evolution
$$ A = e^{-i \Delta H}\ ,$$ and we take the $\Delta \rightarrow 0$ limit.  In this limit the unitary operator expressed in the ordered orthonormal basis is tri-diagonal and closely related to the corresponding expression for the Hamiltonian expressed in the Krylov basis.  
In this section we prove this important link and thus provide a concrete experimental interpretation of spread complexity.  \\ \\ 
\subsection{$|\kappa_1\rangle$}

An explicit expression for $|\kappa_1\rangle$ can be obtained as 
\begin{eqnarray}
|\kappa_1\rangle & = & N\left( A |\kappa_0\rangle - \langle \kappa_0 | A \ignore{| \tilde{K}_0\rangle } |\kappa_0\rangle|\kappa_0\rangle \right) \ , \nonumber \\ 
N & = & \left( \langle \kappa_0| A^\dag A | \kappa_0 \rangle - \langle \kappa_0| A^\dag| \kappa_0 \rangle  \langle \kappa_0| A | \kappa_0 \rangle  \right)^{-\frac{1}{2}} \ .    \nonumber
\end{eqnarray}
Specialising to $A = e^{- i \Delta H}$ and expanding in the $\Delta \rightarrow 0$ limit yields
\begin{eqnarray}
\frac{|\kappa_1 \rangle}{N} & \rightarrow & -i \Delta \left( H |\kappa_0\rangle - \langle \kappa_0 | H | \kappa_0\rangle  |\kappa_0 \rangle \right) \ ,  \nonumber \\
N & \rightarrow & \frac{1}{i \Delta \sqrt{ \langle \kappa_0 | H^2 | \kappa_0 \rangle  - \langle \kappa_0 | H | \kappa_0 \rangle ^2   } }  \ , \nonumber
\end{eqnarray}
which combines to yield the first Krylov basis vector generated by the Hamiltonian $H$ from the reference state $|\kappa_0\rangle$.  

\subsection{$| \kappa_2 \rangle $}

As with $|\kappa_1\rangle$ we can obtain an explicit expression for this vector.  However, it is easier to generalise if we change our approach.  Note that the orthogonality of the vectors 
\begin{equation}
 \langle \kappa_0 | \kappa_2\rangle =    \langle \kappa_0 | \kappa_1\rangle = 0 \ ,
\end{equation}
is independent of $\Delta$.  As such, the orthogonality is also true as a Taylor expansion in $\Delta$.  We have already shown that $|\kappa_0\rangle$ and $|\kappa_1\rangle$ tend to $|K_0\rangle$ and $|K_1\rangle$ in the $\Delta \rightarrow 0$ limit. These provide an (ordered) orthogonal basis for space of states spanned by $\left\{ |K_0\rangle, H|K_0\rangle  \right\}$.  For $|\kappa_2\rangle$ we have that
\begin{eqnarray}
|\kappa_2) &=& c_{2,2} e^{-2i \Delta H} |\kappa_0\rangle + c_{2,1} e^{-i \Delta H}|\kappa_0\rangle + c_{2,0} |\kappa_0\rangle    \nonumber \\
& = & a |\kappa_0\rangle + b \Delta H |\kappa_0\rangle + c \Delta^2 H^2 |\kappa_0\rangle  + o(\Delta^3) \nonumber \ ,
\end{eqnarray}
so that, in the $\Delta \rightarrow 0$ limit, $|\kappa_2)$ becomes a linear combination of $|\kappa_0\rangle$, $H|\kappa_0\rangle$ and $H^2|\kappa_0\rangle$ that's orthogonal to $|K_0\rangle$ and $|K_1\rangle$.  This is, of course, the Krylov vector $|K_2\rangle$.  By induction it can now be shown that same holds for all $|\kappa_n\rangle$ i.e. that in the $\Delta\rightarrow 0$ limit we have
$$ |\kappa_n\rangle \ \rightarrow |K_n\rangle  \ \ \ \forall n \ . $$
In addition, the cost function we have selected in the previous subsection becomes
\begin{equation}
\hat{K} = \sum_{n} n |\kappa_n \rangle \langle \kappa_n | \rightarrow \sum_{n} n |K_n \rangle \langle K_n | \ ,
\end{equation}
which is the usual spread complexity cost operator.  Taken together, this implies that the computational cost we have defined above is equivalent to spread complexity in the $\Delta \rightarrow 0$ limit.  \\ \\
Before proceeding, it is important to stress that what we have presented here is not in contradiction with the no-go theorems that were proven in \cite{Aguilar-Gutierrez:2023nyk}.  Therein the authors demonstrated that Krylov complexity cannot be used to define a notion of distance on the space of accessible operator (or states in the context of spread complexity).  More precisely, taking 
$$ C(|\psi_t\rangle; H, |\psi_r\rangle) $$
to be the spread complexity of the state $|\psi_t\rangle$ with Krylov basis generated from $|\psi_r\rangle$ through the Hamiltonian $H$, the authors demonstrated that the triangle inequality 
\begin{equation}
C( |\psi_c\rangle; H, |\psi_a\rangle    ) \leq C( |\psi_c\rangle; H, |\psi_b\rangle    ) + C( |\psi_b\rangle; H, |\psi_a\rangle    ) 
\end{equation}
is not satisfied.  
The reason that our construction is not in contradiction with this important result this is that our computational circuits are designed to \textbf{always} take a specific reference state as the input state.  As such, the notion of distance between different an intermediate state and the target state cannot appear. It would be interesting to see, however, if the circuit complexity perspective we have proposed here allows for a well-defined metric on the space of target states\footnote{The point is that, once optimized, the sequence of operations that constitute the channels are fixed.  One may then compute the result of passing different input states through these channels.}. 

\section{Computation of Circuit Complexity}

As explained at the end of section (\ref{CircuitComplexitySection}), since the gates we consider are unitary (and not hermitian) we will not obtain a tri-diagonal expression for the unitary expressed in the ordered basis.  As a consequence the usual iterative procedure to find the orthonormal basis (\cite{Lanczos1950AnIM}) is not possible.  In this section we unpack another computational scheme one may apply to compute the circuit complexity. 
\\ \\
The crucial property is that the states $|\kappa_n\rangle$ are orthogonal and involve a total of $n$ terms, see eq. (\ref{KtildeBasis}).  Using this we may write 
\begin{eqnarray}
\sum_{m,m'} c_{a,m} S_{m m'} c^*_{b,m'} & = & N_a \delta_{ab}  \ ,   \nonumber \\
(C S C^\dag)_{ab} & = & N_a \delta_{ab}  \ ,  \nonumber
\end{eqnarray}
where we have defined
\begin{eqnarray}
C_{m,n} & = & c_{m,n} \ ,        \nonumber \\
S_{m,m'} & = & \langle \kappa_0|  (A^\dag)^{m'} A^{m} |\kappa_0\rangle \ .
\label{RMatrix}
\end{eqnarray}
Note that the matrix $C$ is a lower triangular matrix with $1$'s on the diagonal and that the matrix $S$ is positive-definite since\\ $N_a > 0 \ \forall a$. It then follows that the matrix $C$ may be derived from the $LU$-decomposition (or generally an $LDU$-decomposition) of the matrix $S_{m,m'}$.  More precisely, since the matrix $S$ is hermitian and positive-definite, this allows for a Cholesky decomposition.   Explicitly, this gives the expansion coefficients as
\begin{eqnarray}
S & = & L D U     \nonumber \\
\Rightarrow \ \ \ C & = & L^{-1} = (U^\dag)^{-1} \\
N_a &=& (L^{-1} S (L^{-1})^\dag )_{aa} = D_{aa} \ .   \nonumber
\end{eqnarray}
Armed with the solution for the coefficients, we can now compute the ordered orthogonal basis necessary to calculate the circuit complexity.  We denote the circuit complexity as $C\left( |\psi\rangle  ; A, |\kappa_0\rangle \right)$ where $|\psi\rangle$ is the target state, $|\kappa_0\rangle$ the reference state and $A$ the unitary.  We find
\begin{eqnarray}
C\left( |\psi\rangle  ; A, |\kappa_0\rangle \right) & = & \sum_{n} n |\langle \kappa_n| \psi\rangle|^2     \nonumber \\
& = &  \sum_{n} n \frac{\langle \psi | \kappa_n)( \kappa_n | \psi \rangle  }{(\kappa_n | \kappa_n)}    \nonumber \\
& = &  \sum_{n,m,m'} n \frac{\ c_{n,m} c_{n,m'}^*}{N_n} \langle \psi| A^m   | \kappa_0\rangle\langle \kappa_0 |(A^\dag)^{m'} |\psi \rangle   \ ,    \nonumber
\end{eqnarray}
to a general state in the space of target states.  

\subsection{Time-evolution as the unitary gate}

Consider now the case where we want to compute the computational complexity of a state that is the result of a finite transformation generated by $A$.  In the case of time-translations we are interested in computing the complexity of the state $|\psi\rangle = e^{-i t H} |\kappa_0 \rangle$.  In this case we find that 
\begin{equation}
C\left( e^{-i t H} |\kappa_0 \rangle  ; e^{-i \Delta H}, |\kappa_0\rangle \right) = \sum_{n} n  \frac{(C \rho C^\dag)_{nn}}{(C S C^\dag)_{nn}   } \ ,  \label{discreteK}
\end{equation}
where we have defined
\begin{equation}
\rho_{m,m'} = \langle \kappa_0| e^{i (t - m \Delta) H}   | \kappa_0\rangle \langle \kappa_0 |e^{-i(t - m' \Delta) H} |\kappa_0 \rangle    \ ,\label{rhoMatrix}
\end{equation}
and the matrices $S, C$ are given in (\ref{RMatrix}).  At this stage it is important to reflect on this result.  All the entries in $C$, $\rho$ and $S$ depend \textbf{only} on the return amplitude 
\begin{equation}
R(\tau) = \langle \kappa_0 | e^{-i \tau H} |\kappa_0\rangle \ .
\end{equation}
The matrix $S$ is computed by combining the return amplitudes evaluated at a discrete number of points $R(\Delta k)$.  The matrix $C$ and normalisation constants $N_a$ are then computed by means of an $LU$-decomposition of $S$ and thus also only depends on the return amplitude evaluated at these discrete points.  The matrix $\rho$ contains products of the return amplitude evaluated at shifted times.  The expression (\ref{discreteK}) is thus determined \textbf{in full} by specifying the return amplitude.  In particular, no derivatives need to be computed.  

\subsection{An analytic $SU(2)$ example}

\begin{figure}[h]
        \begin{center}
            \includegraphics[width=0.45\textwidth]{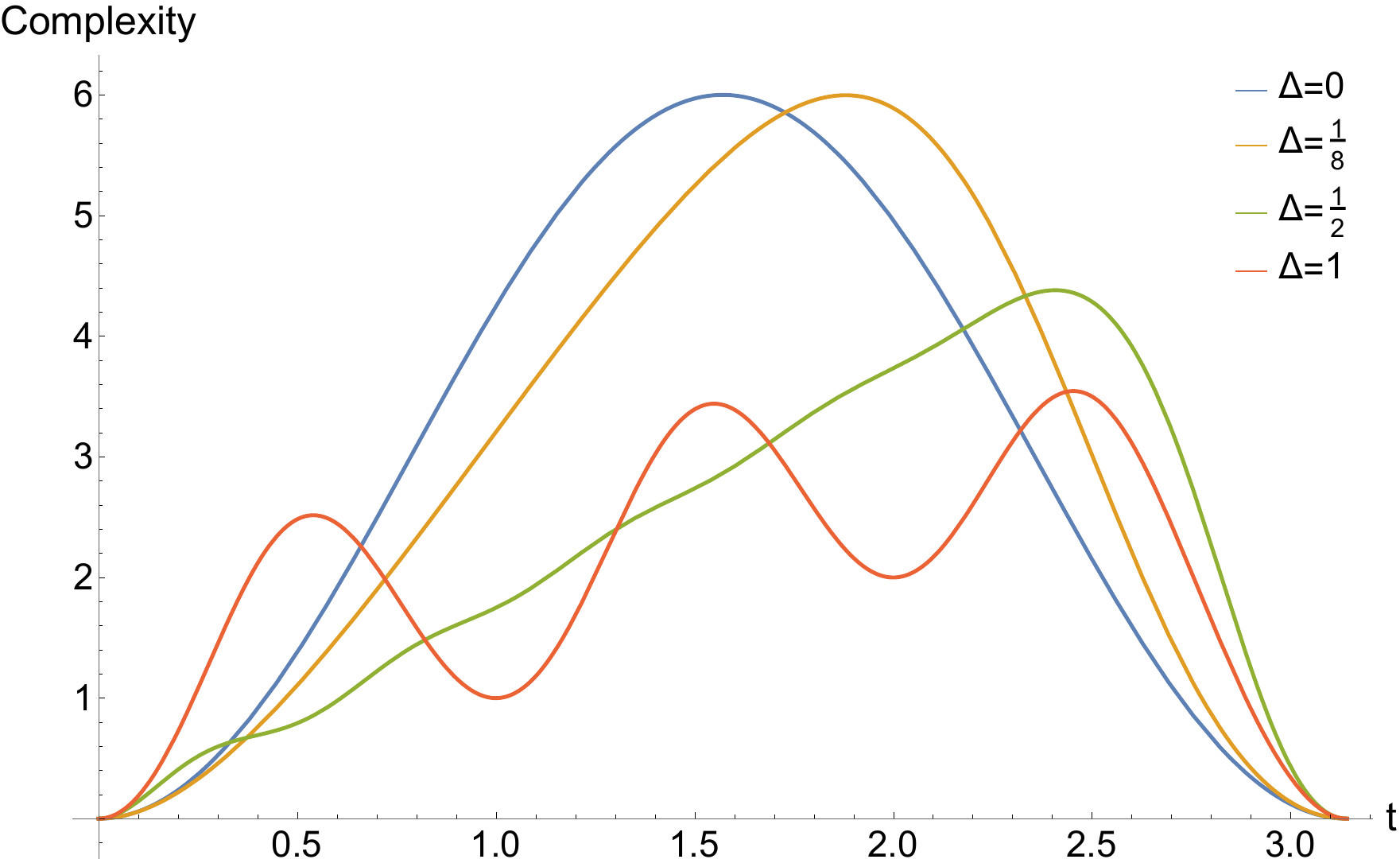}
        \caption{The circuit complexity for the $SU(2)$ Hamiltonian for various values of $\Delta$ and $j=3, \alpha=1$ and $|\kappa_0\rangle = |j,-j\rangle$.  The $\Delta=0$ case is the usual spread complexity.  }
        \label{CircComp}
        \end{center}
\end{figure}

\begin{figure}[h]
        \begin{center}
            \includegraphics[width=0.45\textwidth]{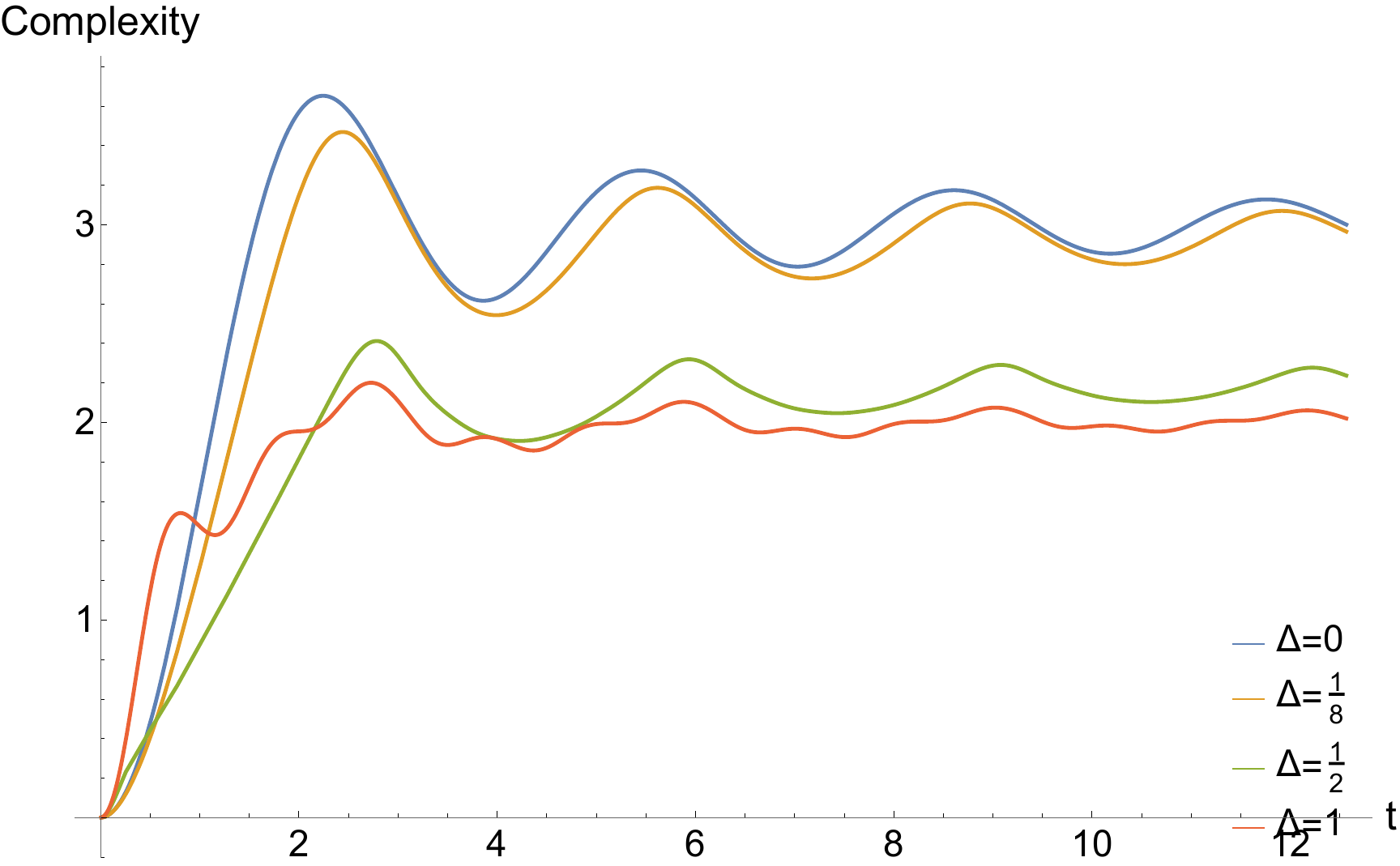}
        \caption{The time-averaged circuit complexity for the $SU(2)$ Hamiltonian for various values of $\Delta$ and $j=3, \alpha=1$ and $|\kappa_0\rangle = |j,-j\rangle$.  The $\Delta=0$ case is the usual time-averaged spread complexity.  }
        \label{AveragedCircuitComp}
        \end{center}
\end{figure}

As an explicit example, consider the following Hamiltonian
\begin{equation}
H = \alpha(J_{+} + J_{-}) \ ,
\end{equation}
constituted of $su(2)$ generators which satisfy
\begin{eqnarray}
\left[ J_{+}, J_{-}  \right] & = & 2 J_0   \ ,\nonumber \\
\left[ J_0, J_{\pm}  \right] & = & \pm J_{\pm} \ .  \nonumber
\end{eqnarray}
The return amplitude, for $|\kappa_0\rangle = |j,-j\rangle$ is given by
\begin{equation}
R(\tau) = \left( \cos(\alpha \tau) \right)^{2j} \ .
\end{equation}
An instructive example is $j=1$.  In this case we obtain the following relevant matrices
\begin{eqnarray}
S & = & \left( \begin{array}{ccc} 1 & \cos^2(\alpha \Delta) & \cos^2(2 \alpha\Delta) \\ \cos^2(\alpha\Delta) & 1 & \cos^2(\alpha \Delta) \\ \cos^2(2\alpha \Delta) & \cos^2(\alpha \Delta) & 1   \end{array}   \right)    \nonumber \\
\Rightarrow C & = & \left( \begin{array}{ccc}  1 & 0 & 0 \\ -\cos^2(\alpha\Delta) & 1 & 0 \\ \frac{1 + 3 \cos(2\alpha\Delta)}{3 + \cos(2\alpha \Delta)}  & -\frac{8 \cos^4(\alpha\Delta)}{3 + \cos(2\alpha \Delta)} & 1     \end{array}   \right)  \ ,   \nonumber
\end{eqnarray}
and
\begin{equation}
\rho_{m m'} = \cos^2(\alpha (t - m\Delta)) \cos^2(\alpha (t - m' \Delta)) \ .  
\end{equation}
Note that the matrices above are all simple functions of $R(\tau)$.  
Using these matrices we can compute the circuit complexity as
\begin{eqnarray}
C(t) & = & 2 \sin^2(\alpha t)  + \frac{(1 + 3 \cos(2 \alpha t))\sin^2(\alpha \Delta)}{3 + \cos(2\alpha\Delta)} \sin^2(\alpha t) \nonumber \\
& & -\frac{2 \sin(2\alpha t)\sin(2\alpha \Delta)}{3 + \cos(2\alpha\Delta)}\sin^2(\alpha t)   \ . \nonumber
\end{eqnarray}
In the limit $\Delta \rightarrow 0$ we recover the known expression for spread complexity for this setup. \\ \\
We can (in principle) obtain expressions for other values of $j$ though these are not particularly insightful.  In Fig. (\ref{CircComp}) we have rather plotted the circuit complexity for $j=3$ and various values of $\Delta$.  \\ \\
As $\Delta$ is decreased the profile clearly approaches the known expression for spread complexity.  A noteworthy feature of this circuit complexity (for $\Delta \neq 0$) is that it is periodic but not symmetric under time-reversal.  The reason for this is simple: our unitary gate performs a small forward time-step with $\Delta$ and there isn't a second gate that can (for the same cost) perform a small backward time-step in $\Delta$.  If included in our analysis this would restore the time-reversal symmetry.  \\ \\
Furthermore, the Krylov basis (obtained for $\Delta =0$) is clearly the basis that minimises the complexity of the time-evolved state at early times, as was proven in \cite{Balasubramanian:2022tpr}.  For finite times this may not be the case.  For example, when $\Delta = 1$ our construction ensures that e.g. the circuit complexity at $t=1$ should have an upper bound of $1$.  As a consequence, the time-averaged complexity may be minimised by a basis generated from some finite time-step.  In Fig. (\ref{AveragedCircuitComp}) we have plotted the corresponding time-averaged complexities to illustrate this. 

\subsection{Connection with Nielsen complexity}
\noindent
A connection between Krylov complexity and geometric (Nielsen) complexity\footnote{See also \cite{Craps:2023ivc} where a relation between Nielsen complexity and the time-average of Krylov complexity has also been established in more general terms} has been previously established for systems whose Hilbert space dynamics are governed by group manifolds associated with Lie algebras (\cite{Caputa:2021sib, Chattopadhyay:2023fob}). In such cases, the Krylov basis can be related to coherent state manifolds generated by Lie group actions on a highest-weight or vacuum state. Specifically, consider a Lie algebra containing standard raising and lowering operators $L_{+}, L_{-}$, with dynamics governed by a Hamiltonian (or Liouvillian) of the form,
\begin{equation}
H\rightarrow{\cal L}=\alpha(L_++L_-)\;,
\end{equation}
and associated displacement operator $D(\xi)=e^{\xi L_+-\bar\xi L_i}$, which generates generalized coherent states via 
\begin{equation}
|z\rangle=D(\xi)|0\rangle\;,
\end{equation}
where $|0\rangle$ is the reference ({\it e.g.} vacuum or highest weight) state and $z$ is a complex parameter labeling the manifold.  The geometry of the resulting state manifold is encoded in the defined by the Fubini-Study metric,
\begin{equation}
ds^2=\frac{1}{\langle z|z\rangle}
\left[\langle dz|dz\rangle-\frac{\langle dz|z\rangle\langle z|dz\rangle}{\langle z|z\rangle }\right]=g_{z\bar z}dzd\bar z\ ,
\end{equation}
which serves as the information metric for the quantum evolution within this family of states. A classical Hamiltonian ${\cal H}(z,\bar z)=\langle z|\hat H|z\rangle$ can be defined on this manifold, with associated Poisson bracket structure
\begin{equation}
\{A,B\}=g^{z\bar z}\left[\frac{\partial A}{\partial  \bar z}\frac{\partial B}{\partial z}-\frac{\partial A}{\partial z}\frac{\partial B}{\partial \bar z} \right]\;,
\end{equation}
where $g^{z\bar{z}}$ is the inverse of the information metric. The equations of motion governing evolution of this manifold are then given by,
\begin{equation}
z'(t)=\{z,{\cal H}(z,\bar z)\}\;,\;\;\bar z'(t)=
\{\bar z,{\cal H}(z,\bar z)\}\ ,
\end{equation}
and define a classical trajectory through the quantum state space. This evolution is identified with geodesic motion in the Nielsen complexity geometry, with an appropriately chosen cost functional. In this framework, the Krylov complexity can be shown to be proportional to the geometric Nielsen complexity computed with the $F_{1}$ cost function,
\begin{equation}
C_K(t) \propto F_1(t) = |\langle z(t) | \delta z(t) \rangle|\ ,
\end{equation}
which measures the tangent vector norm along the path of coherent states in the geometry.  \\ 
Given our earlier results connecting circuit complexity and Krylov complexity, this geometric construction establishes a direct bridge between circuit complexity and Nielsen complexity. In particular, it elucidates how operator growth (as measured by Krylov complexity) manifests as motion through the space of unitaries with minimal cost, according to the chosen Nielsen geometry. This reinforces the interpretation of Krylov complexity as a physically meaningful and computationally tractable proxy for circuit complexity in quantum many-body systems.

\subsection{Possible Applications and Generalizations}

The main contribution of this letter is an explicit physical interpretation of spread complexity.  It is clear that spread complexity may be viewed as circuit complexity with a gate for superposition and another for (infinitesimal) time-evolution.  We do expect, however, that this formulation may also find a useful application under certain circumstances.  \\ \\
The matrices $S$ and $\rho$ involve the return amplitude evaluated at discrete points in time.  For $S$ these are in the vicinity of $t=0$ and for $\rho$ these are in the vicinity the desired value of $t$.  Only the value of the return amplitude is needed and none of its derivatives.  In cases where the late-time return amplitude may be approximated (e.g. by a saddle point approximation) it may be more feasible to compute the circuit complexity than the traditional spread complexity.  Furthermore, for some choices of Hamiltonian and reference state, the expectation value $\langle \kappa_0 | H^m | \kappa_0\rangle$ may diverge for some finite value of $m$.  In this case, though the return amplitude is well-defined, its perturbative expansion in $t$ may not converge.  In such cases the Lanczos algorithm breaks down.  Spread complexity may, however, still be computed as a limiting case of the circuit complexity.  \\ \\
The scheme we have outlined that utilizes a single unitary gate may be generalized to several unitary gates (along with superposition).  This is a fascinating possible future direction which may make contact with multiseed Krylov complexity (\cite{Craps:2024suj}) with an appropriate choice of cost function.  As is the case with spread complexity, this would provide a direct physical interpretation of multi-seed Krylov complexity.  \\ \\
As a final comment, we note that a paper by \cite{Lv:2023jbv} has previously established a connection between Krylov complexity and circuit complexity when dynamical symmetries are present.  The circuit complexity referred to therein is rather different to the one we have used in this paper, the former relating to a continuous manifold of unitary circuits and ours being discrete quantum gates.  Furthermore, the connection we have established here makes no assumptions on the system Hamiltonian and, in particular, does not require the existence of dynamical symmetries.

\section*{Acknowledgements}
\noindent
The work of HN is supported in part by  CNPq grant 304583/2023-5 and FAPESP grant 2019/21281-4.
HN would also like to thank the ICTP-SAIFR for their support through FAPESP grant 2021/14335-0. 
ELG is supported by FAPESP grant 2024/13362-2.
JM and HJRVZ would like to acknowledge support from  the “Quantum Technologies for
Sustainable Development” grant from the National Institute
for Theoretical and Computational Sciences of South Africa
(NITHECS). CB is supported by the Oppenheimer Memorial Trust PhD Fellowship and
the Harry Crossley PhD Fellowship.
\appendix

\section{Channel Designs}\label{Experiment}

\begin{figure}[h]
        \begin{center}
            \includegraphics[width=0.49\textwidth]{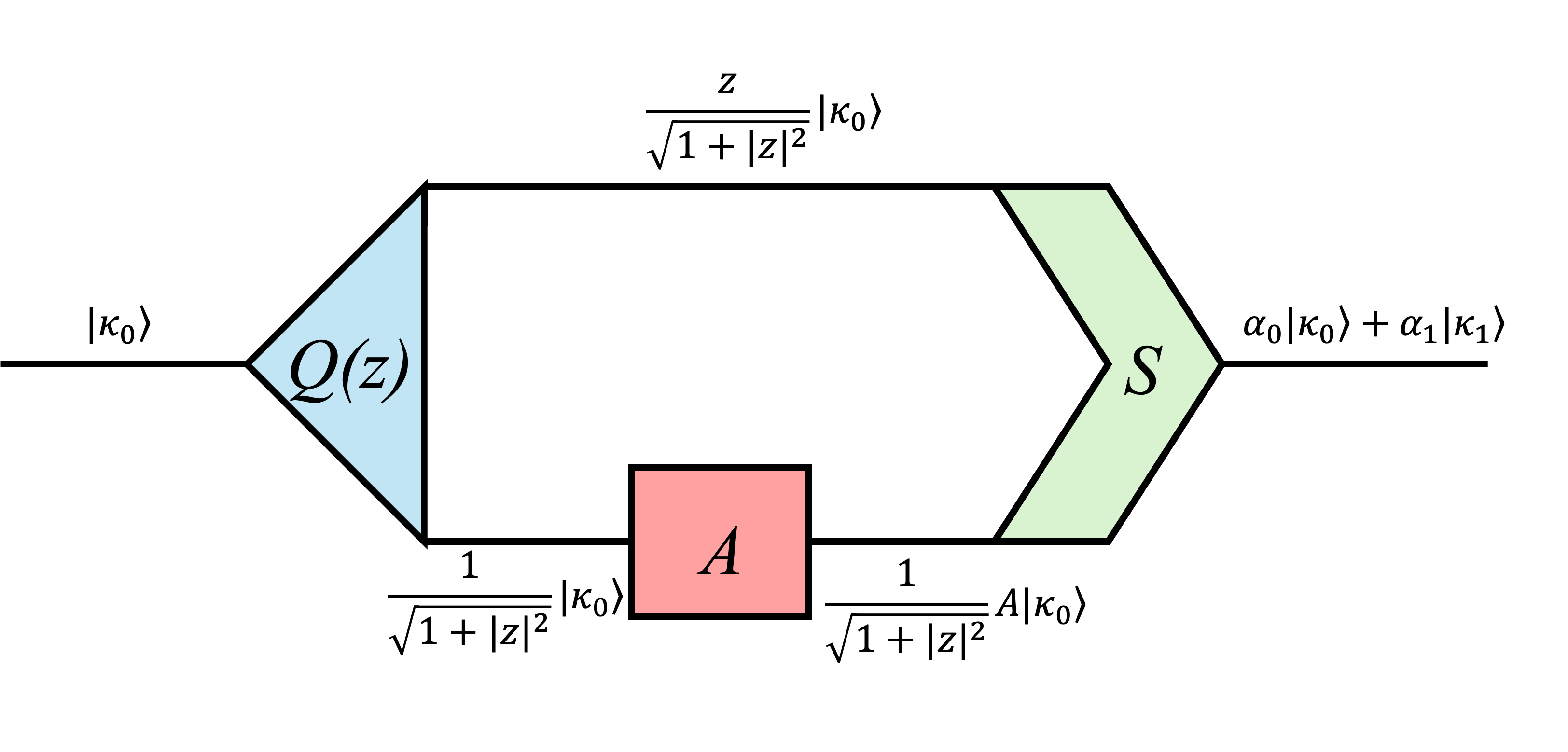}
        \caption{An example design for the channel $C_1$.  }
        \label{fig:C1Design}
        \end{center}
\end{figure}

The operations $Q$, $S$ and $A$ can be combined to generate the circuits of interest in the main text.  We begin by sketching an example design for channel $C_1$ 
\begin{eqnarray}
S_{12} A_1 Q_1(z) |\kappa_0\rangle & = & S_{12} A_1 \left( \begin{array}{c}   \frac{1}{\sqrt{1 + |z|^2}} |\kappa_0\rangle \\  \frac{z}{\sqrt{1 + |z|^2}} |\kappa_0\rangle  \end{array} \right)   \nonumber \\
& = & S_{12} \left( \begin{array}{c}   \frac{1}{\sqrt{1 + |z|^2}} A |\kappa_0\rangle \\  \frac{z}{\sqrt{1 + |z|^2}} |\kappa_0\rangle  \end{array} \right)  \ .    \nonumber
\end{eqnarray}
Before computing the superposition it is useful to decompose the state $A|\kappa_0\rangle$ into orthogonal states as 
$$  A|\kappa_0\rangle  =   \langle \kappa_0| A |\kappa_0\rangle \ |\kappa_0\rangle + \sqrt{1 - |\langle \kappa_0| A |\kappa_0\rangle|^2} |\kappa_1\rangle \ .  $$ 
Now performing the superposition yields
\begin{equation}
S_{12} A_1 Q_1(z) |\kappa_0\rangle
= \frac{\sqrt{1 - |\langle \kappa_0| A |\kappa_0\rangle|^2} |\kappa_1\rangle + \left(  z + \langle \kappa_0 | A |\kappa_0\rangle   \right)|\kappa_0\rangle}{\sqrt{1 + |z|^2 + z \langle \kappa_0 | A^\dag |\kappa_0\rangle  + \bar{z} \langle \kappa_0 | A |\kappa_0\rangle }} \ .
\end{equation}
By varying $z$ it is clear that we can obtain an arbitrary superposition of the states $|\kappa_0\rangle$ and $|\kappa_1\rangle$.  We have sketched this design in Fig. (\ref{fig:C1Design})
\\ \\
Similarly, an example design for channel $C_2$ can be obtained as
\begin{eqnarray}
& & S_{12}S_{23} A_1 Q_{1}(z') A_1 Q_1(z) |\kappa_0\rangle   \nonumber \\
& = & S_{12}S_{23} A_1 Q_{1}(z')  \left( \begin{array}{c}   \frac{1}{\sqrt{1 + |z|^2}} A |\kappa_0\rangle \\  \frac{z}{\sqrt{1 + |z|^2}} |\kappa_0\rangle  \end{array} \right)     \nonumber \\
& = & S_{12}S_{23}  \left( \begin{array}{c} \frac{1}{\sqrt{1 + |z|^2}\sqrt{1 + |z'|^2}} A^2 |\kappa_0\rangle  \\   \frac{z'}{\sqrt{1 + |z|^2}\sqrt{1 + |z'|^2}} A |\kappa_0\rangle \\  \frac{z}{\sqrt{1 + |z|^2}} |\kappa_0\rangle  \end{array} \right)  \ ,   \nonumber
\end{eqnarray} 
where we have expressed the explicit expression for brevity.  By varying the parameters $z'$ and $z$ we can obtain an arbitrary superposition of the states $|\kappa_0\rangle$, $|\kappa_1\rangle$ and $\kappa_2\rangle$.  Example designs for the other channels can be obtained along very similar lines.  \\ \\
A final comment on our circuits is in order.  While the operations $Q$ and $A$ are unitary operations the operation $S$ is not.  A simple way to see this is to note that it is not invertible.  In practice this means that any scheme by which superposition is implemented would not be fully probability-preserving.  A unitary operation may produce both the desired superposed state and another state.  This is true even for multi-slit experiments where any reflected particles are effectively lost and only those passing through the slits contribute to the superposed image.  For the purposes of our discussion, this does not invalidate our circuits since at least a fraction of the input particles can be put into the desired output state.  However, this should realistically affect the assumption of zero cost of superpositions. A more accurate treatment would also assign some computational cost to both $S$ and $Q$.

\section{Optimal Channel design for our chosen cost function}

\label{AppendixOptimal}

Consider a target state that is a superposition 
$$ |\psi\rangle = \sum_{l=1}^N \  \gamma_{l} |\kappa_l\rangle $$
We can substitute this directly into our chosen cost function (\ref{costFunc}) to find
$$ C_{\kappa} = \sum_{k} \sum_{m} k \left| \overline{\gamma_m} \ \frac{ \alpha_{k, m}   }{ \sqrt{ \sum_{l = 0}^k  | \alpha_{k, l}|^2  }   } \right|^2 =  \sum_{k} \sum_{m} k  |\gamma_m|^2 \ \frac{ |\alpha_{k, m}|^2   }{ \sum_{l = 0}^k  | \alpha_{k, l}|^2  }     .    $$   The above only depends on the ratios $ \tilde{\alpha}_{k,m} \equiv \frac{ |\alpha_{k,m}|  }{ |\alpha_{k,k}| }   $.  The extrema of this function are, of course, determined by solving 
$$ \frac{\partial}{\partial \tilde{\alpha}_{k,m}} C_{\kappa} = 0  $$
and one easily finds that, for each $\tilde{\alpha}$ there are two extrema, the relevant one being $\tilde{\alpha}_{k,m} = 0$.  The other extremum, where $\tilde{\alpha}_{k,m} \rightarrow \infty$ stand in contrast to the aim of our channel design, which is that channel $\hat{C}_{k}$ should contain at least $k$ actions of the unitary.   
\\ \\
To test whether these extrama represent the minimum of the of the function, we must compute the Hessian matrix, which turns out to be diagonal
$$ \left. \frac{\partial^2 }{\partial \tilde{\alpha}_{k,m}  \partial \tilde{\alpha}_{k,m}     }   C_{\kappa} \right|_{\tilde{\alpha}_{a,b} = 0}  = 2 k (|\gamma_{m}|^2 - |\gamma_{k}|^2) $$
so that the channel design in the main text is a minimum for (at least) all states that satisfy
\begin{equation}
|\gamma_0| > |\gamma_1| > |\gamma_2| > \cdots > |\gamma_N|     \label{minCondition}
\end{equation}
When describing the early-time evolution of the reference state, this condition is satisfied.  We thus conclude that the optimal channel design is given by setting all the $\tilde{\alpha}_{a,b} = 0$ which yields
$$ \hat{C}_n = |\kappa_n\rangle \langle \kappa_0| $$
as used in the main text.  \\ \\  An alternative perspective is that the different channel designs are different choices of basis for the set of target states.  The choice of basis $|\kappa_n\rangle$ i.e. the sequence of orthogonal states generated by $A$ and (cost-free) superpositions is optimal as basis, at least for states satisfying (\ref{minCondition}).   Like the Krylov basis is the natural basis to use to describe the spread of the time-evolved reference state since it is the basis that minimises the (early-time) complexity of the time-evolved reference state or, more generally, any state that has the expansion $$  |\psi(x)\rangle = \sum_{ n   } c_n x^n H^n |K_0\rangle$$   the basis $|\kappa_n\rangle$ is the basis that minimises the complexity for small $x$ for any state that permits the expansion
$$|\phi(x)\rangle = \sum_{ m   } c'_m x^m A^m |\kappa_0\rangle. $$In this sense the basis $|\kappa_n\rangle$ is the natural basis to use for the circuit setup we employ.

\

\bibliographystyle{elsarticle-harv} 
\bibliography{complexityRefs}

\end{document}